\documentclass[12pt]{article}
\catcode`\@=11
\def\@normalsize{\@setsize\normalsize{15pt}\xiipt\@xiipt
\abovedisplayskip 14pt plus3pt minus3pt%
\belowdisplayskip \abovedisplayskip
\abovedisplayshortskip  \z@ plus3pt%
\belowdisplayshortskip  7pt plus3.5pt minus0pt}
\def\small{\@setsize\small{13.6pt}\xipt\@xipt
\abovedisplayskip 13pt plus3pt minus3pt%
\belowdisplayskip \abovedisplayskip
\abovedisplayshortskip  \z@ plus3pt%
\belowdisplayshortskip  7pt plus3.5pt minus0pt
\def\@listi{\parsep 4.5pt plus 2pt minus 1pt
            \itemsep \parsep
            \topsep 9pt plus 3pt minus 3pt}}
\def\underline#1{\relax\ifmmode\@@underline#1\else
        $\@@underline{\hbox{#1}}$\relax\fi}
\@twosidetrue \relax \catcode`@=12
\catcode`\@=11

\def\section{\@startsection{section}{1}{\z@}{3.5ex plus 1ex minus
   .2ex}{2.3ex plus .2ex}{\large\bf}}

\def\ps@headings{\def\@oddfoot{}\def\@evenfoot{}
\def\@oddhead{\hbox{}\hfill
        \makebox[.5\textwidth]{\raggedright\ignorespaces --\thepage{}--
        \hfill }}
\def\@evenhead{\@oddhead}
\def\subsectionmark##1{\markboth{##1}{}}
} \ps@headings \catcode`\@=12 \relax

\def\figcap{\section*{Figure Captions\markboth
        {FIGURECAPTIONS}{FIGURECAPTIONS}}\list
        {Fig. \arabic{enumi}:\hfill}{\settowidth\labelwidth{Fig. 999:}
        \leftmargin\labelwidth
        \advance\leftmargin\labelsep\usecounter{enumi}}}
 \relax
\def\tablecap{\section*{Table Captions\markboth
        {TABLECAPTIONS}{TABLECAPTIONS}}\list
        {Table \arabic{enumi}:\hfill}{\settowidth\labelwidth{Table 999:}
        \leftmargin\labelwidth
        \advance\leftmargin\labelsep\usecounter{enumi}}}
 \relax
\def\reflist{\section*{References\markboth
        {REFLIST}{REFLIST}}\list
        {[\arabic{enumi}]\hfill}{\settowidth\labelwidth{[999]}
        \leftmargin\labelwidth
        \advance\leftmargin\labelsep\usecounter{enumi}}}
 \relax
\catcode`\@=11
\def\marginnote#1{}
\newcount\hour
\newcount\minute
\newtoks\amorpm
\hour=\time\divide\hour by60 \minute=\time{\multiply\hour by60
\global\advance\minute by- \hour}
\edef\standardtime{{\ifnum\hour<12 \global\amorpm={am}%
    \else\global\amorpm={pm}\advance\hour by-12 \fi
    \ifnum\hour=0 \hour=12 \fi
    \number\hour:\ifnum\minute<100\fi\number\minute\the\amorpm}}
\edef\militarytime{\number\hour:\ifnum\minute<100\fi\number\minute}

\def\draftlabel#1{{\@bsphack\if@filesw {\let\thepage\relax
  \xdef\@gtempa{\write\@auxout{\string
    \newlabel{#1}{{\@currentlabel}{\thepage}}}}}\@gtempa
    \if@nobreak \ifvmode\nobreak\fi\fi\fi\@esphack}
     \gdef\@eqnlabel{#1}}
\def\@eqnlabel{}
\def\@vacuum{}
\def\draftmarginnote#1{\marginpar{\raggedright\scriptsize\tt#1}}
\def\draft{\oddsidemargin -.5truein
        \def\@oddfoot{\sl preliminary draft \hfil
        \rm\thepage\hfil\sl\today\quad\militarytime}
        \let\@evenfoot\@oddfoot \overfullrule 3pt
        \let\label=\draftlabel
        \let\marginnote=\draftmarginnote
\def\@eqnnum{(\theequation)\rlap{\kern\marginparsep\tt\@eqnlabel}%
\global\let\@eqnlabel\@vacuum}  }
\def\preprint{\twocolumn\sloppy\flushbottom\parindent 1em
        \leftmargini 2em\leftmarginv .5em\leftmarginvi .5em
        \oddsidemargin -.5in    \evensidemargin -.5in
        \columnsep 15mm \footheight 0pt
        \textwidth 250mmin      \topmargin  -.4in
        \headheight 12pt \topskip .4in
        \textheight 175mm
        \footskip 0pt
\def\@oddhead{\thepage\hfil\addtocounter{page}{1}\thepage}
        \let\@evenhead\@oddhead \def\@oddfoot{} \def\@evenfoot{}
}
\def\titlepage{\@restonecolfalse\if@twocolumn\@restonecoltrue\onecolumn
     \else \newpage \fi \thispagestyle{empty}\c@page\z@
        \def\thefootnote{\fnsymbol{footnote}} }
\def\endtitlepage{\if@restonecol\twocolumn \else  \fi
        \def\thefootnote{\arabic{footnote}}
        \setcounter{footnote}{0}}  
\catcode`@=12 \relax

\def\ps@headings{\def\@oddfoot{}\def\@evenfoot{}
\def\@oddhead{\hbox{}\hfill
        \makebox[.5\textwidth]{\raggedright\ignorespaces --\thepage{}--
        \hfill }}
\def\@evenhead{\@oddhead}
\def\subsectionmark##1{\markboth{##1}{}}
} \ps@headings \relax
\newcommand{\newc}{\newcommand}
\newc{\ra}{\rightarrow}
\newc{\lra}{\leftrightarrow}
\newc{\beq}{\begin{equation}}
\newc{\be}{\begin{equation}}
\newc{\eeq}{\end{equation}}
\newc{\ee}{\end{equation}}
\newc{\bea}{\begin{eqnarray}}
\newc{\eea}{\end{eqnarray}}

\newc{\ome}{\omega}
\newc{\ba}{\begin{eqnarray}}
 \newc{\ea}{\end{eqnarray}}

\newcommand{\lb}{{\varepsilon}}
\begin{document}
\def\firstpage#1#2#3#4#5#6{
\begin{titlepage}
\nopagebreak
\title{\begin{flushright}
        \vspace*{-0.8in}
{ \normalsize  hep-ph/yymmddd\\
May 2005 \\
}
\end{flushright}
\vfill {#3}}
\author{\large #4 \\[1.0cm] #5}
\maketitle \vskip -7mm \nopagebreak
\begin{abstract}
{\noindent #6}
\end{abstract}
\vfill
\begin{flushleft}
\rule{16.1cm}{0.2mm}\\[-3mm]

\end{flushleft}
\thispagestyle{empty}
\end{titlepage}}

\def\simlt{\stackrel{<}{{}_\sim}}
\def\simgt{\stackrel{>}{{}_\sim}}
\date{}
\firstpage{3118}{IC/95/34} {\large\bf Majorana Neutrino Masses
from
Flavor Symmetries\footnote{Talk given in the "Workshop on Recent Advances in Particle Physics
and Cosmology", Thessaloniki 2005}} {A. Psallidas}
{\normalsize\sl Theoretical Physics Division, The University of
Ioannina,
GR-45110 Ioannina, Greece\\[2.5mm]
 }
{ In this talk we discuss the implications of the Minimal
Supersymmetric Standard Model augmented by a single $U(1)$
anomalous family symmetry for neutrino masses and mixing angles.
The left-handed neutrino states are provided with Majorana masses
through a dimension-five operator in the absence of right handed
neutrino components. Assuming symmetric lepton mass matrices, the
model predicts inverse hierarchical neutrino mass spectrum,
$\theta_{13}=0$ and  large mixing  while at the same time it
provides acceptable mass matrices for the charged fermions. }

\vskip 3truecm

\newpage

\section{Introduction}

 Enlarging the gauge symmetry of the Standard Model is a common
solution to some of its problems. A natural candidate would be an
additional $U(1)$ family symmetry that is broken at some high
scale $M$, a scenario proposed some time ago for the explanation
of the charged fermion mass hierarchy~\cite{Froggatt:1979nt,
ir,Dent:2004dn} as well as  for the realization of the
quark-lepton complementarity (QLC)~\cite{Petcov:1993rk} more
recently ~\cite{  Raidal:2004iw, Antusch:2005ca} . This is
motivated by the experience from string model building which has
shown that a natural step towards this simplification is to assume
the existence of $U(1)$ symmetries which distinguish the various
families.

Experimental facts~\cite{rdata} suggest that the Yukawa couplings
related to neutrino masses are highly suppressed compared to those
of quarks and charged leptons while their mixing is much larger
than that of the quark sector. Therefore, exploring whether the
neutrino oscillation data could be interpreted in the context of
an extension of the Standard Model gauge symmetry is an
interesting possibility. We will use only the minimal fermion
spectrum  of the Minimal Supersymmetric Standard Model (MSSM)
without introducing right handed neutrinos~\cite{Leontaris:2004rd}
. Thus, we will provide Majorana masses for all three neutrinos
from the  lepton number violating operator \cite{Weinberg:sa},
which has the form \beq \frac{(\bar{L}^c H) (L H)}{M}
\label{neff}\eeq where $M$ stands for a large scale that will turn
out to be of the order $10^{13-14}$ GeV. This scale is quite low
to be identified with the GUT  or the string scale in the context
of the heterotic string theory, however, it is compatible with the
effective gravity scale   in theories with large extra
 dimensions  obtained in the context of Type I string models.

 In this talk, we  explore the possibility  that neutrino masses and mixing
can be interpreted with the help of an additional anomalous $U(1)$
family symmetry which at the same time is responsible for the
generation of charge fermion mass hierarchy. This symmetry could
be anomalous and anomaly cancellation is assumed to happen in the
context of a fundamental theory valid above the scale $M$. We show
that in a generic model an additional abelian symmetry can account
for atmospheric data and predicts $\theta_{13}=0$. We also show
how secondary effects possibly arising from additional singlet(s)
or some alternative mechanism, as supersymmetry breaking, can
under certain assumptions render the model compatible with all
recent experimental
 data. We finally derive explicit charge assignments that reproduce the above results.

\section{\label{secc}Description of the Model}

We  consider the MSSM with gauge symmetry $G_{SM}=SU(3)\times
SU(2)_L\times U(1)_Y$ as an effective field theory below a scale
$M$ of a fundamental theory. In the context of  the $G_{SM}$
symmetry, all  gauge invariant Yukawa terms relevant to quark and
charged lepton masses appearing at the tree-level superpotential
are \ba {\cal W} &=& y_{ij}^u Q_i U^c_j H_2 + y_{ij}^dQ_i D^c_j
H_1
               +y_{ij}^e L_{i} E^c_j H_1.
\label{sup1} \ea In the case of models constructed in the
framework of string theory, there are explicit examples where the
MSSM fields are charged under (at least) one additional abelian
anomalous ($U(1)_X$) factor that prevents terms not invariant
under this symmetry from appearing in (\ref{sup1}). Usually, the
appearance of the additional $U(1)_X$ symmetry is accompanied by
at least a pair of MSSM singlets ($\Phi$, $\bar{\Phi}$) with
opposite $U(1)_X$-charges. $\Phi$ and  $\bar{\Phi}$ can acquire
vevs leading to the breaking of the extra abelian symmetry.

Assuming natural values of the Yukawa couplings $\lambda_{ij}$ in
(\ref{sup1}) (i.e., order one), and taking into account the
observed low energy hierarchy of the fermion mass spectrum, we
infer that only  couplings associated with the third generation
should remain invariant at tree-level. Mass terms for the lighter
fermions are to be generated from higher order non-renormalizable
superpotential couplings. Such higher order invariants are formed
by adding to the non-invariant tree-level coupling an appropriate
number of $U(1)_X$-charged singlet fields which compensate the
excess of the $U(1)_X$-charge. In the case supersymmetric models,
the magnitudes of the singlet vevs $\langle\Phi\rangle$ and
$\langle\bar\Phi\rangle$ are related by the $D$-flatness
conditions of the superpotential, while perturbative
considerations require that the vevs for the singlet fields are
about one order of magnitude below the effective theory scale $M$
scale, therefore lighter generations couplings will be suppressed
by powers of $\lambda$, $\bar{\lambda}$ where \ba
\lambda=\frac{\langle\Phi\rangle}{M},\,
 \bar{\lambda}=\displaystyle\frac{\langle{\bar\Phi}\rangle}{M}\label{exp}
\ea Introducing the generic charge $U(1)_X$-charge assignments of
Table~\ref{uxa}, the charges of the entries of  the corresponding
mass matrices are \ba C^u_{ij}=q_i+u_j\ ,\  C^d_{ij}=q_i+d_j\ ,\
C^e_{ij}=\ell_i+e_j. \label{cd} \ea Restricting  the analysis to
the investigation of symmetric fermion mass matrices
\begin{table}
\begin{center}
\begin{tabular}{|cc|cc|}
\hline
Fermion&Charge&Higgs&Charge\\
\hline
$Q_i(3,2,\frac 16)$&$q_i$&$H_1(1,2,-\frac 12)$&$h_1$\\
$D^c_i(\bar 3,1,\frac 13)$&$d_i$&$H_2(1,2,\frac 12)$&$h_2$\\
$U^c_i(\bar 3,1,-\frac 23)$&$u_i$&&\\
$L_i(1,2,-\frac 12)$&$\ell_i$&$\Phi(1,1,0)$&$+1$\\
$E^c_i(1,1,1)$&$e_i$&$\bar\Phi(1,1,0)$&$-1$\\
\hline
\end{tabular}
\end{center}
\caption{\label{uxa}$U(1)_X$ charge assignments for MSSM fields.
The $U(1)_X$ charges of the two extra singlet  fields $\Phi$ and
$\bar\Phi$, are taken to be $+1$ and $-1$ respectively. }
\end{table}
we obtain the following constraints $ q_i+u_j=q_j+u_i $, $
q_i+d_j=q_j+d_i $ , $ \ell_i+e_j=\ell_j+e_i $ . Moreover, the
requirement that the third generation mass couplings appear at
tree-level imposes the additional constraints $q_3+u_3+h_2=0$,
$q_3+d_3+h_1=0$, $ \ell_3+e_3+h_1 = 0$. Since in our
configurations the top, bottom and $\tau$--Yukawa couplings are
equal at the high scale $M$, up to order one coefficients, the
difference between the top mass $(m_t)$ and the bottom mass
($m_b$) must arise mainly from a large Higgs vev ratio $\tan\beta
=\frac{v_2}{v_1}\gg 1$.

The general form of the superpotential couplings contributing to
the fermion  mass matrices has been studied in
~\cite{Leontaris:2004rd}. Here, we will  concentrate on the
neutrinos.  These are massless at tree-level, however, the
non-renormalizable mass term (\ref{neff}) leads directly to a
light Majorana mass matrix involving only the left handed
components $\nu_{Lj}$. Therefore, defining $\varepsilon^{k}=
\lambda^{k} \ \mbox{if} \ k=[k]<0$, $\varepsilon^{k} =
\bar{\lambda}^{k} \ \mbox{if} \ k=[k]>0$ and $\varepsilon^{k} =0 \
\mbox{if} \ k\ne[k]$ (where $[k]$ stands for the integer part of
$k$ ) the mass term takes the form \ba
  {\cal W}_{n.r.}^{(2)}= \frac{\zeta_{\nu}^{a \beta} }{M}
  \varepsilon^{C^\nu_{ij}}(\bar{L_{a}^{c}}^{i} H_{2}^{j} \epsilon_{ji})
  (H_{2}^{l} L_{\beta}^{k}  \epsilon_{lk}) &
  \equiv& {\zeta_{\nu}^{a \beta} }\,\varepsilon^{C^\nu_{ij}}\,
 \frac{v_2^2}{M}\,\bar\nu_{La}^c\,\nu_{L\beta}\label{d5}
  \ea
with $v_2=\langle{H_2}\rangle \approx O(m_W)$ and
${C^\nu_{ij}}=2h_2+\ell_i+\ell_j$.

 For the quarks we
impose  \be q_1-q_3=\frac{n}{2}\ , \ q_2-q_3 =\frac{m}{2}\ \
\mbox{\rm where}\ m+n\ne0,\ m,n=\pm1,\pm2,\dots\label{qi} \ee
Details for the quark sector can be found in
~\cite{Leontaris:1999wf}.

For the leptons we define the parameters $2n'=l_1-l_3$ and
$2m'=l_2-l_3$, where $m',n'$ are integers and the associated
$U(1)_X$--charge matrix takes the form \ba
C_{e}=\left(\begin{array}{ccc}
n'&\frac{m'+n'}{2}&\frac{n'}{2}\\
\frac{m'+n'}{2}&{m'}&\frac{m'}{2}\\
\frac{n'}{2}&\frac{m'}{2}&0\\
\end{array}\right)
\label{lep1} \ea The zero charge in the position 33 of the above
charge-matrices is due to the fact that we demand the appearance
of the corresponding Yukawa couplings at the tree-level
superpotential. For the remaining entries, a proper power of the
appropriate expansion parameter is needed.

We can re-express the generic fermion charges of Table 1 in terms
of the new parameters which we choose to be $m,n$, $m',n'$ that
appear in the quark and charged lepton matrices and
$q_3,\ell_3,h_2,h_1$. The resulting assignments are presented in
Table 2.
\begin{table}
\centering
\begin{tabular}{|l|c|c|c|}
\hline
field&\multicolumn{3}{c|}{generation}\\
\hline
&1&2&3\\
\hline
$Q$&$\frac{n}{2}+q_3$&$\frac{m}{2}+q_3$&$q_3$\\
$U^c$&$\frac{n}{2}-q_3-h_2$&$\frac{m}{2}-q_3-h_2$&$-h_2-q_3$\\
$D^c$&$\frac{n}{2}-q_3-h_1$&$\frac{m}{2}-q_3-h_1$&$-h_1-q_3$\\
$L$&$\frac{n'}{2}+l_3$&
$\frac{m'}{2}+l_3$&$l_3$\\
$E^c$&$\frac{n'}{2}-l_3-
h_1$&$\frac{m'}{2}-l_3-h_1$&$-h_1-l_3$\\
\hline
\multicolumn{4}{|c|}{Higgs}\\
\hline
$H_1$&$h_1$&$H_2$&$h_2$\\
\hline
\end{tabular}
\caption{Fermion $U(1)_X$ charge assignments after introducing the
integer parameters $m,n$ and $m',n'$ that appear in the quark and
charge lepton matrices respectively.} \label{tsol1}
\end{table}

The $U(1)_X$-charge entries for  the light Majorana neutrino mass
matrix take the form \ba C_{\nu}= \left(\begin{array}{ccc}
n'+{\cal A}&\frac{m'+n'}{2}+{\cal A}&\frac{n'}{2}+{\cal A}\\
\frac{m'+n'}{2}+{\cal A}&{m'}+{\cal A}&\frac{m'}{2}+{\cal A}\\
\frac{n'}{2}+{\cal A}&\frac{m'}{2}+{\cal A}&{\cal A}\\
\end{array}\right)
\ea where we have introduced the new parameter $
  {\cal A}= 2 (l_3+h_2)
  $.
We observe that the neutrino $U(1)_X$--charge entries differ from
the corresponding charged leptonic entries by the constant ${\cal
A}$ \ba
  C^{\nu}_{ij} = C^{e}_{ij} + {\cal A}\label{diffNL}
\ea

\section{Neutrino Masses and Mixing}

In this section we search for explicit ${U(1)}_X$ charge
assignments for MSSM particles that provide phenomenologically
acceptable mass textures for all MSSM fermions and in particular
for neutrinos. The basic structure of the mass matrices and mixing
angles which meet the phenomenological requirements can be
obtained without referring to a set of particular
$U(1)_X$-charges. Explicit examples with sets of charges for all
fermion and Higgs fields will be
 given in the end of this section. Before we present viable cases, we should note that our procedure
exhibits here the basic structure of the mass matrices and mixing.
The most striking feature, is that the extension of the $G_{SM}$
symmetry to include an $U(1)_X$ anomalous factor can reproduce the
correct hierarchy of all fermion fields while at the same time the
recent neutrino oscillation data are interpreted to a good
approximation  by a lepton mixing matrix involving two  mixing
angles, one originating from the charged leptonic matrix matrix
and the second by the light Majorana mass matrix. However, at this
level of analysis the value of the non-vanishing coefficients of
the Yukawa superpotential terms are unknown, since their
calculation requires a detailed knowledge of the fundamental
theory above the scale $M$ (possibly string theory). Hence, in the
present analysis, we restrict ourselves in the description
 of the general characteristics of the theory, which are nevertheless very interesting.

We first note that in our framework the leptonic matrices depend
on $m', n'$. We can  fix  the parameters $m,n$, so that a correct
hierarchical quark mass spectrum is
obtained~\cite{Leontaris:2004rd} .  The lepton sector can be then
worked out independently, choosing appropriate values for the two
additional parameters $m'$ and $n'$.

In  order to obtain a viable set of lepton mass matrices and
mixing, a systematic search shows that the charge parameters
$m',n'$ should be $n' = \mbox{odd} \ ,\ m' = \mbox{even}$. Under
this choice  the charged lepton
 mass matrix takes the form
\ba
   M_e = m_0^e\left(
  \begin{array}{ccc}
  \delta\,\lb^{n'} &  0 & 0 \\
  0  &
   \lb^{m'} & \alpha\,\lb^\frac{m'}{2}  \\
   0 &
      \alpha\,\lb^\frac{m'}{2}  & 1 \\
  \end{array}
  \right)
  \label{palep3}
\ea where we have explicitly introduced two (out of three)
order-one parameters $a$ and $\delta$ that account for the Yukawa
couplings and renormalization effects.

Turning to the neutrino sector the Majorana neutrino mass matrix
takes the form
 \beq
  M_{\nu}^0 =m_0^{\nu}\,\left(
  \begin{array}{ccc}
  0 &  -\lb^{ \frac{m'+n'}{2}+{\cal A}}
  & \zeta\,\lb^{ \frac{n'}{2}+{\cal A}} \\
   -\lb^{\frac{m'+n'}{2}+{\cal A}}&
   0 &  0\\
   \zeta\,\lb^{\frac{n'}{2}+{\cal A}} &
      0&  0 \\
  \end{array}
  \right)
  \label{panet3}
  \eeq
where $\zeta$ stands for an order one coefficient. This mass
matrix can be diagonalised by a unitary matrix $V_{\nu}(\omega)$,
where $\tan\ome =\zeta\, \lb^{-m'/2}$, and can lead to bimaximal
mixing in the case that the two mass matrix elements are equal
~\cite{Frampton:2004ud}.

The leptonic mixing matrix $U_l^0 = V_l^\dagger(\phi)
V_{\nu}(\omega)$ is given by \ba U_l^0&= &\left(\begin{array}{ccc}
 -\frac{1}{\sqrt{2}}& \frac{1}{\sqrt{2}}& 0 \\
-\frac{\cos(\phi+\ome)}{\sqrt{2}}&-\frac{\cos(\phi+\ome)}{\sqrt{2}}&\sin(\phi+\ome)\\
\frac{\sin(\phi+\ome)}{\sqrt{2}}&\frac{\sin(\phi+\ome)}{\sqrt{2}}&\cos(\phi+\ome)\\
\end{array}\right)\label{bimax}
\ea
 where $\tan(2\phi) = 2\,a\,\lb^{\frac{m'}{2}}/(1-\lb^{m'})$ while for the mass square differences
\ba \Delta m^2_{atm}=\Delta m_{23}^2={(m_0^{\nu})}^2\,{\lb^{2{\cal
A} +{m'+n'}{}}}({ 1+\zeta^2\lb^{-m'} })\ ,\ \Delta
m^2_{\odot}=\Delta m_{12}^2=0 \label{neig} \ea
 The above results exhibit a number of interesting properties of
the model, that are worth mentioning at this point. We first
observe that the model predicts an inverted neutrino mass
hierarchy. We further point out that the $U(1)_X$ symmetry implies
large mixing effects in the neutrino mass matrix, in contrast to
the situation of the charged fermion sector where the mixing is
small.  Moreover, at this level of approximation,   a $zero$-entry
for the element $U_{13}$ is predicted in the mixing matrix.  The
rest of the elements are determined by two angles, $\phi$ arising
from the charged lepton mass matrix diagonalisation and
 $\ome$ arising from the neutrino mass matrix.

 Working out specific cases  we aim to find explicit sets of
$U(1)_X$ charges which interpret the neutrino data in the context
of the above scenario. For the specific solutions  we have set
$m'$ even and $n'$ odd. Then, from the formulae of Table
(\ref{tsol1}), we find that the leptons have fractional
$U(1)_X$-charges  of the form $\frac{2 k+1}4$, with $k$ integer.
\begin{table}
\centering
\begin{tabular}{|l|c|c|c|}
\hline
\multicolumn{4}{|c|}{Solution A}\\
\hline
field&\multicolumn{3}{c|}{generation}\\
\hline
&1&2&3\\
\hline
$Q$&$4$&$2$&$0$\\
$D^c$&$2$&$0$&$-2$\\
$U^c$&$4$&$2$&$0$\\
$L$&$\frac{9}{4}$&$-\frac{1}{4}$&$-\frac{5}{4}$\\
$E^c$&$\frac{11}{4}$&$\frac{1}{4}$&$-\frac{3}{4}$\\
\hline
\multicolumn{4}{|c|}{Higgs}\\
\hline
$H_1$&$2$&$H_2$&$0$\\
\hline
\multicolumn{4}{|c|}{Singlets}\\
\hline
$\Phi$&$1$&$\bar\Phi$&$-1$\\
\hline
\end{tabular}
\begin{tabular}{|l|c|c|c|}
\hline
\multicolumn{4}{|c|}{Solution B}\\
\hline
field&\multicolumn{3}{c|}{generation}\\
\hline
&1&2&3\\
\hline
$Q$&$4$&$2$&$0$\\
$D^c$&$1$&$-1$&$-3$\\
$U^c$&$4$&$2$&$0$\\
$L$&$\frac{9}{4}$&$-\frac{1}{4}$&$-\frac{5}{4}$\\
$E^c$&$\frac{7}{4}$&$-\frac{3}{4}$&$-\frac{7}{4}$\\
\hline
\multicolumn{4}{|c|}{Higgs}\\
\hline
$H_1$&$3$&$H_2$&$0$\\
\hline
\multicolumn{4}{|c|}{Singlets}\\
\hline
$\Phi$&$1$&$\bar\Phi$&$-1$\\
\hline
\end{tabular}
\caption{Examples of $U(1)_X$ charges which lead to the neutrino
mass matrix structure discussed in the text.} \label{charges}
\end{table}

Choosing for example, the values $m=4,n=8,m'=2,n'=7, h_1 =2,h_2 =0
, A =-\frac{5}{2}$ we obtain the charge assignments of solution A
of Table \ref{charges} and the following fermion mass matrices for
the quarks \ba
 M_{u,d} \sim  m_0^{u,d}\left(\begin{array}{ccc}
\lb^8&\lb^6&\lb^4\\
\lb^6&\lb^4&\lb^2\\
\lb^4&\lb^2&1\\
\end{array}\right).
\ea (which is the texture discussed in~\cite{Leontaris:1999wf}),
the charged leptons \ba M_e \sim m_0^e\left(\begin{array}{ccc}
\delta\,\lb^{7}&0&0\\
0&\lb^2&a\,\lb\\
0&a\,\lb&1\\
\end{array}\right)\label{lex}
\ea and the neutrinos \ba M_\nu^0 \sim
m_0^\nu\left(\begin{array}{ccc}
0&-\lb^{2}&\zeta\,\lb\\
-\lb^{2}&0&0\\
\zeta\,\lb&0&0\\
\end{array}\right)\label{lexx}.
\ea

Charged lepton masses can be fit  within a range of the mass
matrix  parameters in (\ref{lex}). For example,  choosing
$\lb\sim0.28$, $\alpha\sim-1.3$ and $\delta\sim2$, the correct
mass spectrum is obtained. Atmospheric neutrino oscillation
mass-squared difference is then reproduced for $M\sim 5\times
10^{13} GeV$ modulo order one coefficients. This scale is quite
low to be identified with the string scale in heterotic
constructions, it is however compatible with type I superstring
models where the string scale is tight to the Planck scale. Other
configurations of additional $U(1)_X$-charges are also possible
since the mass matrices under consideration do not depend on the
parameters $q_3,\ell_3$. For example choosing solution B of Table
\ref{charges} we obtain the same mass matrices as in solution A
considered above.

As already noted however,  at this level of analysis,  the
neutrino mass splitting between the first and second generation
does not appear because the two eigenstates are degenerate.
Moreover, the solar neutrino mixing angle is maximal, a situation
disfavored by recent data. This discrepancy can be lifted however,
if additional non-zero entries are generated by hierarchically
smaller effects.   We find it interesting that two additional
entries, for example 11 and 23,  smaller than the entries 12 and
13 already present at this level, would be sufficient to bring the
final form of the neutrino matrix to an acceptable two-zero
texture mass matrix ~\cite{Frampton:2002qc}, that provides the
necessary mass splitting  and interpret accurately the
experimental data. To show that this is indeed the case, let us
assume that, after the inclusion of these effects and in the basis
where the charged lepton mass matrix is diagonal, the neutrino
mass matrix takes the form \ba
  M_{\nu} =m_0^{\nu}\,\left(
  \begin{array}{ccc}
  2 x &  -\cos\,\bar\ome & \sin\,\bar\ome  \\
   -\cos\,\bar\ome& 0 &  2 y\\
   \sin\,\bar\ome &  2 y&  0 \\
  \end{array}
  \right)
  \label{mncor}
  \ea
  where $\bar\omega =\ome+\phi$.

The neutrino mass-squared differences have a ratio which depends
on a different $x,y$ linear combination, $(x-y\sin(2\bar\ome)$,
\ba \frac{\Delta m_{12}^2}{\Delta m_{23}^2}&=&\frac{4
(x-y\sin(2\bar\ome))}{1-2 (x-y\sin(2\bar\ome))}\label{ratio} \ea
while the mixing angles are analogously
corrected~\cite{Leontaris:2004rd} \beq \tan\theta_{23} \ \approx \
\tan\bar\omega \ , \   \tan\theta_{13} \ \approx \
2y\cos(2\bar\omega) \ , \  \tan\theta_{12} \ \approx \ 1-(x+y
\sin(2\bar\omega))  \eeq

Using the experimental data we find that experimentally acceptable
$\tan\theta_{12}$ values can be satisfied for  $x\approx
[0.10-0.24]$ and $y\approx [0.10-0.22]$,  assuming $\bar\omega$ to
be maximal. We remark that these values in a wide portion of the
acceptable range, are sufficiently smaller that the order one 12-
and 13-neutrino mass matrix entries and thus our approximation is
consistent.

\section{Conclusions}

In this talk, we have presented a simple extension of the Minimal
Supersymmetric Standard model by an anomalous $U(1)_X$ symmetry
broken at some high scale $M$ and attempted to interpret the
recent neutrino experimental data using just the left-handed
neutrino components. Assuming symmetric mass matrices and that the
third generation of up, down quarks and charged fermions acquire
masses at tree-level, we derive the general charge assignments for
MSSM fermions and examine their implications for the Majoranna
neutrino mass matrix resulting from the dimension 5 operator
$(LH)^2/M$. We find that the model leads naturally to inverted
mass hierarchy for neutrinos, $\theta_{13}=0$ and maximal
atmospheric mixing for  $M\sim10^{13-14}GeV$. At this level the
absolute masses of the lightest eigenstates are equal and solar
neutrino mixing turns out to be also maximal. We show that higher
appropriate order corrections  lift the mass degeneracy and the
solar neutrino data can be accurately described. We derive
explicit fermion ${U(1)}_X$ charge assignments that realize the
above scenario.

{\bf Ackmowledgements}. {\it This research was funded by the
program "Heraklitos" of the Operational Program for Education and
Initial Vocational Training of the Hellenic Ministry of Education
under the 3rd Community Support Framework and the European Social
Fund.}

\end{document}